\documentclass[a4paper]{jpconf}
\usepackage{graphicx}
\begin{document}
\title{75 years of double beta decay: yesterday, today and tomorrow}

\author{A S Barabash}

\address{Institute of Theoretical and Experimental Physics, 
B. Cheremushkinskaya 25, 117218 Moscow, Russia}

\ead{barabash@itep.ru}

\begin{abstract}
In this report I will briefly review the motivation and history of double beta decay search since 
the first consideration of two neutrino process (2$\beta(2\nu)$) by Maria Goeppert-Mayer in 1935. The first experiments 
on search for double beta decay in the late of 1940's and beginning of 1950's are considered. It is underlined that for the first 
time the 2$\beta(2\nu)$ decay has been registered in geochemical experiment with $^{130}$Te in 1950. In direct (counter) 
experiment this type of decay for the first time has been registered in $^{82}$Se by Michael Moe's group in 1987. Now two 
neutrino double beta decay has been recorded for 10 nuclei ($^{48}$Ca, $^{76}$Ge, $^{82}$Se, $^{96}$Zr, $^{100}$Mo, $^{116}$Cd, 
$^{128}$Te, $^{130}$Te, $^{150}$Nd, $^{238}$U). In addition, the 2$\beta(2\nu)$ decay of $^{100}$Mo and $^{150}$Nd 
to the 0$^+_1$ excited state of the daughter nucleus has been observed and the ECEC(2$\nu$) process in $^{130}$Ba was observed too. 
As to neutrinoless double beta decay (2$\beta(0\nu)$) this process has not yet been registered. In the review results 
of the most sensitive experiments (Heidelberg-Moscow, IGEX, CUORICINO, NEMO-3) are discussed and conservative 
upper limits on effective Majorana neutrino mass and the coupling constant of the Majoron to the neutrino are established 
as $\langle m_{\nu} \rangle < 0.75$ eV and 
$\langle g_{ee} \rangle < 1.9 \cdot 10^{-4}$, respectively. 
     The next-generation experiments, where the mass of the isotopes being studied will be 
as grand as 100 to 1000 kg, are discussed. These experiments will have started within a few years. In all probability, 
they will make it possible to reach the sensitivity to the neutrino mass at a level of 0.01 to 0.1 eV.

\end{abstract}

\section{Introduction}

The current interest in neutrinoless double beta decay
is that the existence of this process is
closely related to the following fundamental aspects of 
particle
physics \cite{KLA98,FAE01,VER02}: (i) lepton-number
non-conservation, (ii) the
presence of a neutrino mass and its origin, (iii) the existence of
right-handed
currents in electroweak interactions, (iv) the existence of the
Majoron, (v) the
structure of the Higgs sector, (vi) supersymmetry, (vii) the existence
of leptoquarks,
(viii) the existence of a heavy sterile neutrino, and (ix) the
existence of a
composite neutrino.

All of these issues are beyond the standard model of electroweak
interaction,
therefore the detection
of $0\nu\beta\beta$ decay would imply the discovery of new physics.
Of course, now
interest in this process is caused primarily by the problem of a
neutrino
mass. If $0\nu\beta\beta$ decay is discovered, then according to
current
thinking, this will automatically mean that the rest mass of at
least
one neutrino flavor is nonzero and is of Majorana origin.

Interest in neutrinoless double-beta decay has seen a significant renewal in 
recent years after evidence for neutrino oscillations was obtained from the 
results of atmospheric, solar, reactor and accelerator  neutrino 
experiments (see, for example, the discussions in \cite{VAL06,BIL06,MOH06}). 
These results are impressive proof that neutrinos have a non-zero mass. However,
the experiments studying neutrino oscillations are not sensitive to the nature
of the neutrino mass (Dirac or Majorana) and provide no information on the 
absolute scale of the neutrino masses, since such experiments are sensitive 
only to the difference of the masses, $\Delta m^2$. The detection and study 
of $0\nu\beta\beta$ decay may clarify the following problems of neutrino 
physics (see discussions in \cite{PAS03,MOH05,PAS06}):
 (i) lepton number non-conservation, (ii) neutrino nature: 
whether the neutrino is a Dirac or a Majorana particle, (iii) absolute neutrino
 mass scale (a measurement or a limit on $m_1$), (iv) the type of neutrino 
mass hierarchy (normal, inverted, or quasidegenerate), (v) CP violation in 
the lepton sector (measurement of the Majorana CP-violating phases).

\section{Yesterday}

The double beta decay problem arose practically immediately after the appearance 
of W. Pauli's neutrino hypothesis 
in 1930 and the development of $\beta$-decay theory by E. Fermi in 1933. 
In 1935 M. Goeppert-Mayer identified for the first time 
the possibility of two neutrino double beta decay, 
in which there is a transformation of an (A, Z) 
nucleus to an (A, Z+2) nucleus that is accompanied by the emission of two electrons 
and two anti-neutrinos \cite{GOE35}:

\begin{equation}
(A,Z) \rightarrow (A,Z+2) + 2e^{-} + 2\tilde \nu                     
\end{equation}

It was demonstrated theoretically by E. Majorana in 1937 \cite{MAJ37} that if one 
allows the existence of only one type of neutrino, which has no antiparticle 
(i.e. $\nu \equiv\tilde\nu$) , then the conclusions of $\beta$-decay theory 
are not changed. In this case one deals with a Majorana neutrino. 
In 1939 W. Furry introduced a scheme of neutrinoless double beta decay 
through the virtual state of intermediate nucleus \cite{FUR39}: 

\begin{equation}
(A,Z) \rightarrow (A,Z+2) + 2e^{-}                                 
\end{equation}

The first experiment to search for $2\beta$-decay was done in 1948 using Geiger counters. 
In this experiment a half-life limit for $^{124}$Sn was established, 
$T_{1/2} > 3\cdot 10^{15}$
 y \cite{FIR48}. During the period from 1948 to 1965 $\sim$ 20 experiments were carried out with a sensitivity 
to the half-life on the level of $\sim 10^{16}-10^{19}$ y (see reviews \cite{LAZ66,HAX84}). The $2\beta$-decay 
was thought to have been "discovered" a few times, but each time it was not confirmed by new 
(more sensitive) measurements. The exception was the geochemical experiment, 
in which two neutrino double beta decay of $^{130}$Te was really detected in 1950 \cite{ING50}. 
     
At the end of the 1960s and beginning of 1970s significant progress in the 
sensitivity of double beta decay experiments was realized. E. Fiorini et al. 
carried out experiments with Ge(Li) detectors and established a limit on 
neutrinoless double beta decay of $^{76}$Ge, $T_{1/2} > 5\cdot 10^{21}$ y \cite{FIO73}. 
Experiments with $^{48}$Ca and $^{82}$Se using streamer chamber with a magnetic field and 
plastic scintillators were done by C. Wu's group and led to impressive limits of 
$ > 2\cdot 10^{21}$ y \cite{BAR70} and $ > 3.1\cdot 10^{21}$ y \cite{CLE75} respectively. During 
these years many sensitive geochemical experiments were done and 
$2\nu\beta\beta$ decay of $^{130}$Te, $^{128}$Te and $^{82}$Se was 
detected (see reviews \cite{KIR83,HAX84,MAN86}). 
      
In 1981 a new type of neutrinoless decay with Majoron emission was introduced \cite{GEO81}:

\begin{equation}
(A,Z) \rightarrow (A,Z+2) + 2e^{-} + \chi^{0}                     
\end{equation}

The important achievements in the 1980s were connected with the first evidence of 
two neutrino double beta decay in direct counting experiments. This was 
done by M. Moe's group for $^{82}$Se using a TPC ($T_{1/2}=1.1^{+0.8}_{-0.3}\cdot10^{20}$ y) \cite{ELL87}. 
There was also the first use 
of semiconductor detectors made of enriched Ge in the ITEP-ErPI experiment \cite{VAS90}.
     
During the 1990s the two neutrino decay process was detected in many experiments 
for different nuclei (see \cite{TRE02,BAR10a}), two neutrino decay to an excited state 
of the daughter nucleus was also detected \cite{BAR95}. In addition, the sensitivity to $0\nu\beta\beta$ 
decay in experiments with $^{76}$Ge (Hidelberg-Moscow \cite{KLA01} and IGEX \cite{AAL02}) 
was increased up to $\sim 10^{25}$ y.  
     
Since 2002 the progress in double beta decay searches has been 
connected mainly with the two experiments, NEMO-3 \cite{ARN05,ARN05a,ARN06,ARN07,ARG09,ARG09a,BAR10} 
and CUORICINO \cite{ARN05b,ARN08,PAV10}.           
The basic historical marks of 75 years study of this process are presented in Tables 1 and 2 (from \cite{BAR10c}).

\begin{table*}[h]
\caption{Main "milestones" in double beta decay search.}
\begin{tabular}{|c|c|c|}
\hline
Date & Event & Remarks \\
\hline
1935 & The idea of 2$\beta2(\nu)$ decay   & M. Goeppert-Mayer \cite{GOE35} \\
 & has been formulated &  \\
 & & \\
1939 & The idea of 2$\beta0(\nu)$ decay   & W.H. Furry \cite{FUR39} \\
 & has been formulated &  \\
& & \\
1948 & The first 2$\beta$ decay experiment  & E.L. Fireman \cite{FIR48}; (Geiger counters and 25 g of \\
 & has been realized &   enriched $^{124}$Sn were used) \\
& & \\
1950 & {\bf The first observation of 2$\beta2(\nu)$}  & {\bf  M.G. Inghram, and J.H. Reynolds \cite{ING50} } \\
& {\bf decay has been done} & {\bf (geochem. experiment with $^{130}$Te); } \\ 
& & $T_{1/2} \approx 1.4\cdot10^{21}$ y \\
1966 & The first counter experiment   & E. Mateosian, and M. Goldhaber \cite{MAT66}  \\
 & with sensitivity higher than $10^{20}$ y & ("detector=source", 11.4 g of enriched $^{48}$Ca);  \\
& has been realized & $T_{1/2}(0\nu) > 2\cdot10^{20}$ y\\
&  &  \\
1967 & The first experiment with   & E. Fiorini et al. \cite{FIO67} (17 cm$^3$ Ge(Li) detector \\
 & semiconductor Ge detector  & on see level); $T_{1/2}(0\nu) > 3\cdot10^{20}$ y \\
& has been realized &  \\
&  & \\
1967 & The observation of 2$\beta(2\nu)$ decay  & T. Kirsten et al. \cite{KIR67} (geochemical experiment); \\
 & of $^{82}$Se has been done & $T_{1/2} \approx 0.6\cdot10^{20}$ y \\
& & \\
1967- & The first counter experiment  & R.K. Bardin, P.J. Gollon, J.D. Ullman, and \\
 1970 & with sensitivity higher than $10^{21}$ y & C.S. Wu \cite{BAR67,BAR70} (strimmer chamber+    \\
 & has been realized & scintillation counters); $T_{1/2}(0\nu$;$^{48}$Ca) $> 2\cdot10^{21}$ y,  \\
 &  & $T_{1/2}(2\nu$;$^{48}$Ca)$ > 3.6\cdot10^{19}$ y \\
& & \\
1973 & The sensitive counter experiment  & E. Fiorini et al. \cite{FIO73} (68 cm$^3$ Ge(Li) detector  \\
 & with $^{76}$Ge has been realized & at 4200 m w.e. depth); $T_{1/2}(0\nu$) $> 5\cdot10^{21}$ y \\
& & \\
1975 & The sensitive counter experiment   &	B.T. Cleveland et al. \cite{CLE75} (streamer chamber +  \\
 & with $^{82}$Se has been realized & scint. counters); $T_{1/2}(0\nu$; $^{82}$Se) $> 3.1\cdot10^{21}$ y \\
& & \\
1980- & The idea of 2$\beta$ decay with Majoron  & Singlet \cite{CHI81}, doublet \cite{AUL82} and triplet \cite{GEL81,GEO81} \\
1981 & emission has been formulated & Majoron has been introduced \\ 
& & \\
1982 & J. Schechter and J.W.F. Valle  & J. Schechter, and J.W.F. Valle \cite{SCH82} (the   \\
 & theorem is formulated  & occurrence of 2$\beta(0\nu)$ decay implies that neutrinos  \\
&  & are Majorana particles with nonzero mass) \\
&  &  \\
1984 & The program to develop low   & E. Fiorini, and T.O. Niinikoski \cite{FIO84}  \\
  & temperature detectors for double  & \\
& beta decay search has been & \\
& formulated  &  \\
1985 & The fundamental theoretical   & M. Doi, T. Kotani, and E. Takasugi \cite{DOI85} (the  \\
  & investigation of double beta decay  & main  formulas for probability of decay, energy   \\
 & has been done & and angular electron spectra have been obtained) \\
& & \\

\hline

\end{tabular}
\end{table*}

\begin{table*}[h]
\caption{Main "milestones" in double beta decay search (continuation of Table 1).}
\begin{tabular}{|c|c|c|}
\hline
Date & Event & Remarks \\
\hline
1986 & The $g_{pp}$ parameter (characterize  & P. Vogel, and M.R. Zirnbauer \cite{VOG86} (within frame- \\
 & the particle-particle interaction   &  works of QRPA models the satisfactory \\
& in nuclei) of QRPA model has & agreement between theoretical and experimental \\
& been introduced & $T_{1/2}(2\nu)$ values for the first time has been observed) \\
&  &  \\
{\bf 1987} & {\bf The first observation of 2$\nu$  } & {\bf M. Moe et al. \cite{ELL87} (TPC with $^{82}$Se;} \\
 & {\bf decay in counter experiment }   & {\bf $T_{1/2}(2\nu)=1.1^{+0.8}_{-0.3}\cdot10^{20}$ y}) \\
& {\bf has been done }& \\
1987- & The first counter experiment   & D.O. Caldwell et al. \cite{CAL91} (8 detectors from natural   \\
1989 & with sensitivity higher than & Ge with full weight 7.2 kg); $T_{1/2}(0\nu) > 1.2\cdot10^{24}$ y \\
&  $10^{24}$ y has been done  &  \\
1987- & The first semiconductor detector  & ITEP-ErPI Collaboration \cite{VAS88,VAS90} (2 detectors \\
1990  & made of enriched germanium   & from enriched Ge with full weight $\sim$ 1.1 kg).\\
 & (86\% of $^{76}$Ge) has been   & In 1990 $T_{1/2}(0\nu) > 1.3\cdot10^{24}$ y  \\
 & started to work. & and $T_{1/2}(2\nu) = (0.9 \pm 0.1)\cdot10^{21}$ y were obtained  \\
& & \\
&  &  \\ 
1991 & The first observation of 2$\nu$ decay   & A.S. Barabash et al. \cite{BAR95} (low background HPGe \\
 & to the excited state of daughter & detector, 1 kg of $^{100}$Mo, $^{100}$Mo-$^{100}$Ru($0^+_1$;1130 keV)  \\
 & nucleus has been done & transition; $T_{1/2}=6.1^{+1.8}_{-1.1}\cdot10^{20}$ y )\\
& & \\
&  &  \\
1990- & The experiments with   & H. Ejiri et al. \cite{EJI91,EJI95}. $2\beta(2\nu)$ decay observation  \\
 1998 & ELEGANT-V detector & in $^{100}$Mo and $^{116}$Cd \\
& & \\
&  &  \\
1991- & The experiments with NEMO-2 & NEMO-2 Collaboration \cite{DAS95,ARN96,ARN98,ARN99}. Study   \\
1997 & detector  & of $2\beta(2\nu)$ decay ($^{100}$Mo, $^{116}$Cd, $^{82}$Se and    \\
 &  & $^{96}$Zr) with registration of all parameters of \\
& & the decay \\
&  &  \\
1991- & The IGEX experiment & Measurements with 6.5 kg of enriched $^{76}$Ge;  \\
1999 &  & $T_{1/2}(0\nu) > 1.57\cdot10^{25}$ y \cite{AAL02} \\
& & \\
1990- & The Heidelberg-Moscow  & Measurements with 11 kg of enriched $^{76}$Ge \cite{KLA01}; \\
2003 & experiment & $T_{1/2}(0\nu) > 1.9\cdot10^{25}$ y,  \\
 &  & $T_{1/2}(2\nu)=1.74 \pm 0.01(stat)^{+0.18}_{-0.16}(syst)\cdot10^{21}$ y \\
& & \\
&  &  \\
2001 & First observation of ECEC(2$\nu$) & Geochemical experiment with $^{130}$Ba, \\
  &  & $T_{1/2} = (2.2 \pm 0.5)\cdot10^{21}$ y  \cite{MES01}\\
& & \\
2002- & NEMO-3 experiment & NEMO-3 Collaboration \cite{ARN05,ARN05a,ARN06,ARN07,ARG09,ARG09a,BAR10};  \\
2010 &  & $T_{1/2}(0\nu$;$^{100}$Mo)$ > 1.1\cdot10^{24}$ y. Observation and \\
 &  & precise investigation of $2\beta(2\nu)$ decay for 7 isotopes\\
 &  &   ($^{48}$Ca, $^{82}$Se, $^{96}$Zr, $^{100}$Mo, $^{116}$Cd, $^{130}$Te, $^{150}$Nd) \\
&  &  \\
2003- & CUORICINO experiment & CUORICINO Collaboration \cite{ARN05b,ARN08,PAV10}; \\
2008 &  & $T_{1/2}(0\nu$;$^{130}$Te)$ > 2.8\cdot10^{24}$ y \\
\hline
\end{tabular}
\end{table*}

\section{Today} 

\subsection{Two neutrino double beta decay}

As discussed above this decay was first recorded in 1950 in a geochemical experiment
with $^{130}$Te \cite{ING50}. In 1967,
it was also found for $^{82}$Se \cite{KIR67}.
Attempts to observe this decay in a direct measurement employing
counters were unsuccessful for
a long time. Only in 1987 could M. Moe, who used a time-projection
chamber (TPC), observe $2\beta(2\nu)$ decay
in $^{82}$Se
for the first time \cite{ELL87}. Within the next few years,
experiments employing counters were able to detect
$2\beta(2\nu)$
decay in many nuclei. In $^{100}$Mo \cite{BAR95,BAR99,BRA01}, and
$^{150}$Nd \cite{BAR04a}
$2\beta(2\nu)$ decay to the $0^{+}$ excited state of the
daughter nucleus was also recorded. The
$2\beta(2\nu)$ decay of $^{238}$U was detected in a radiochemical
experiment \cite{TUR91}, and in a geochemical experiment for the
first time the ECEC process was detected in $^{130}$Ba
\cite{MES01}. Table 3 displays the present-day averaged and
recommended values of
$T_{1/2}$(2$\nu$) from \cite{BAR10a}. At present, experiments devoted to detecting
$2\nu\beta\beta$ decay are approaching a level
where it is insufficient just to record the
decay. It is necessary to
measure numerous parameters of this process to a high precision 
(half-life value, energy sum spectrum, single electron energy spectrum and angular 
distribution).
Tracking detectors that are able to record both
the energy of each electron and the angle at which they diverge
are the most appropriate instruments for
solving this problem. Current tracking NEMO-3 experiment is measuring all parameters of double 
beta decay for seven different nuclei ($^{48}$Ca, $^{82}$Se, $^{96}$Zr, $^{100}$Mo, $^{116}$Cd, 
$^{130}$Te, and $^{150}$Nd) \cite{ARN05,ARN05a,ARN06,ARN07,ARG09,ARG09a,BAR10}.

\begin{table}[h]
\caption{Average and recommended $T_{1/2}(2\nu)$ values (from
\cite{BAR10a}).}
\begin{tabular}{|c|c|}
\hline
Isotope & $T_{1/2}(2\nu)$ \\
\hline
$^{48}$Ca & $4.4^{+0.6}_{-0.5}\cdot10^{19}$ \\
$^{76}$Ge & $(1.5 \pm 0.1)\cdot10^{21}$ \\
$^{82}$Se & $(0.92 \pm 0.07)\cdot10^{20}$ \\
$^{96}$Zr & $(2.3 \pm 0.2)\cdot10^{19}$ \\
$^{100}$Mo & $(7.1 \pm 0.4)\cdot10^{18}$ \\
$^{100}$Mo-$^{100}$Ru$(0^{+}_{1})$ & $(5.9^{+0.8}_{-0.6})\cdot10^{20}$ \\
$^{116}$Cd & $(2.8 \pm 0.2)\cdot10^{19}$\\
$^{128}$Te & $(1.9 \pm 0.4)\cdot10^{24}$ \\
$^{130}$Te & $(6.8^{+1.2}_{-1.1})\cdot10^{20}$ \\
$^{150}$Nd & $(8.2 \pm 0.9)\cdot10^{18}$ \\
$^{150}$Nd-$^{150}$Sm$(0^{+}_{1})$ & $1.33^{+0.45}_{-0.26}\cdot10^{20}$
\\
$^{238}$U & $(2.0 \pm 0.6)\cdot10^{21}$  \\
$^{130}$Ba; ECEC(2$\nu$) & $(2.2 \pm 0.5)\cdot10^{21}$  \\
\hline
\end{tabular}
\end{table}

\begin{table*}[h]
\caption{Best present results on $2\beta(0\nu)$ decay (limits at
90\% C.L.). $^{*)}$ See footnote 1; $^{**)}$ NME from \cite{CAU07} is used;
$^{***)}$ conservative limit from \cite{BER02} is presented}
\begin{tabular}{|c|c|c|c|c|}
\hline
Isotope & $T_{1/2}$, y & $\langle m_{\nu} \rangle$, eV & $\langle
m_{\nu} \rangle$, eV & Experiment \\
& & \cite{KOR07,KOR07a,SIM08,BAR09,CHA08} & \cite{CAU08}  \\
\hline
$^{76}$Ge & $>1.9\cdot10^{25}$ & $<0.22-0.41$ & $<0.69$ & HM
\cite{KLA01} \\
& $\simeq 1.2\cdot10^{25}$(?)$^{*)}$ & $\simeq 0.28-0.52(?)$$^{*)}$ & $\simeq
0.87(?)$$^{*)}$ & Part of HM \cite{KLA04} \\
& $\simeq 2.2\cdot10^{25}$(?)$^{*)}$ & $\simeq 0.21-0.38(?)$$^{*)}$ & $\simeq
0.64(?)$$^{*)}$ & Part of HM \cite{KLA06} \\
& $>1.6\cdot10^{25}$ & $<0.24-0.44$ & $<0.75$ & IGEX
\cite{AAL02} \\
\hline
$^{130}$Te & $>2.8\cdot10^{24}$ & $<0.35-0.59$ & $<0.77$ &
CUORICINO \cite{PAV10} \\
$^{100}$Mo & $>1.1\cdot10^{24}$ & $<0.45-0.93$ & $ - $ & NEMO-
3 \cite{BAR10} \\
$^{136}$Xe & $>4.5\cdot10^{23}$$^{***)}$ & $<1.41-2.67$ & $<2.2$ & DAMA
\cite{BER02} \\
$^{82}$Se & $>3.6\cdot10^{23}$ & $<1.89-1.61$ & $<2.3$ & NEMO-3
\cite{BAR10} \\
$^{116}$Cd & $>1.7\cdot10^{23}$ & $<1.45-2.76$ & $<1.8$$^{**)}$ &
SOLOTVINO \cite{DAN03} \\
\hline
\end{tabular}
\end{table*}

\subsection{Neutrinoless double beta decay}

In contrast to two-neutrino decay, neutrinoless double-beta decay
has not yet been observed
\footnote{The possible
exception is the result with $^{76}$Ge, published by a fraction of
the  Heidelberg-Moscow
Collaboration (see
Table 4). First time the "positive" result was mentioned in \cite{KLA01a}. 
The Moscow part of the
Collaboration does not agree with this conclusion \cite{BAK03} and
there are others who are critical
of this result \cite{AAL02a,ZDE02,STR05}. Thus, at the present time, this
"positive" result is not accepted by the
"2$\beta$ decay community" and it has to be checked.}, 
although it is easier to detect it. In
this case, one seeks, in the
experimental spectrum, a peak of energy equal to the double beta
transition energy and of width determined
by the detector's resolution.

The constraints on the existence of
$0\nu\beta\beta$ decay are presented in Table 4 for the nuclei
for which the best sensitivity has been reached. In calculating constraints
on $\langle m_{\nu} \rangle$, the
nuclear matrix elements from \cite{KOR07,KOR07a,SIM08,BAR09,CHA08} were used (3-d column). It
is advisable to employ the calculations from
these studies, because the calculations are the most thorough and
take into account the most recent theoretical
achievements. In the papers \cite{KOR07,KOR07a,SIM08} $g_{pp}$ values ($g_{pp}$ is 
parameter of the QRPA theory) were fixed using experimental 
half-life values for $2\nu$ decay and then
NME(0$\nu$) were calculated. 
In column four, limits on $\langle m_{\nu} \rangle$,
which were obtained using the NMEs
from a recent Shell Model (SM) calculations \cite{CAU08}, are presented (for $^{116}$Cd NME 
from \cite{CAU07} is used). 

From Table 4 using NME values 
from \cite{KOR07,KOR07a,SIM08,CAU08,BAR09,CHA08}, the limits on 
$\langle m_{\nu} \rangle$ for $^{130}$Te are comparable 
with the $^{76}$Ge results. Now one cannot 
select any experiment as the best one. 
The assemblage of sensitive experiments
for different nuclei permits one to increase the reliability of the limit 
on $\langle m_{\nu} \rangle$. Present conservative limit can be set as 0.75 eV.

\subsection{Neutrinoless double beta decay with Majoron emission}

Table 5 displays the best present-day constraints for an "ordinary"
 Majoron (n = 1).
The "nonstandard" models of the Majoron were experimentally tested
in \cite{GUN97} for $^{76}$Ge
and in \cite{ARN00} for $^{100}$Mo,
$^{116}$Cd, $^{82}$Se, and $^{96}$Zr. Constraints on the decay
modes involving the emission of two Majorons were also
obtained for $^{100}$Mo \cite{TAN93}, $^{116}$Cd \cite{DAN03}, and
$^{130}$Te \cite{ARN03}. In a recent NEMO Collaboration papers \cite{ARN06,ARG09,ARG09a}, 
new results for these processes in $^{100}$Mo, $^{82}$Se, $^{150}$Nd and $^{96}$Zr 
were obtained with the NEMO-3 detector. 
Table 6 gives the best experimental
constraints on decays accompanied by the emission of one or two
Majorons (for n = 2, 3, and 7).
Hence at the present time only limits on double beta decay with
Majoron emission have been obtained (see table 5 and 6).
A conservative present limit on the coupling constant of ordinary 
Majoron to the
neutrino is $\langle g_{ee} \rangle < 1.9 \cdot 10^{-4}$.

\begin{table*}
\caption{Best present limits on $0\nu\chi^{0}\beta\beta$ decay
(ordinary Majoron) at 90\% C.L. The NME from the 
following works were used, 3-d column: \cite{KOR07,KOR07a,SIM08,BAR09,CHA08}, 4-th 
column: \cite{CAU08}. $^{*)}$ Conservative limit from \cite{BER02} is presented;
$^{**)}$ NME from \cite{CAU07} is used.}
\begin{tabular*}{\textwidth}{l@{\extracolsep{\fill}}|c|c|c|}
\hline
Isotope ($E_{2\beta}$, keV) & $T_{1/2}$, y & $\langle g_{ee} \rangle$, \cite{KOR07,KOR07a,SIM08,BAR09,CHA08}
 & $\langle
g_{ee} \rangle$, \cite{CAU08} \\  \\
\hline
$^{76}$Ge (2039) & $>6.4\cdot10^{22}$ \cite{KLA01} & $<(0.54-1.44)\cdot10^{-
4}$ & $<2.4\cdot10^{-4}$ \\
$^{82}$Se (2995) & $>1.5\cdot10^{22}$ \cite{ARN06} & $<(0.58-
1.19)\cdot10^{-4}$ & $<1.9\cdot10^{-4}$ \\
$^{100}$Mo (3034) & $>2.7\cdot10^{22}$ \cite{ARN06} & $<(0.35-
0.85)\cdot10^{-4}$ & - \\
$^{116}$Cd (2805) & $>8\cdot10^{21}$ \cite{DAN03} & $<(0.79-2.56)\cdot10^{-
4}$ & $<1.7\cdot10^{-4}$$^{**)}$ \\
$^{128}$Te (867) & $>1.5\cdot10^{24}$(geochem)\cite{MAN91,BAR10a} & $<(0.63-
1)\cdot10^{-4}$ & $<1.4\cdot10^{-4}$ \\
$^{136}$Xe (2458) & $>1.6\cdot10^{22*)}$ \cite{BER02} & $<(1.51-3.54)\cdot10^{-
4}$ & $<2.9\cdot10^{-4}$ \\
\hline
\end{tabular*}
\end{table*}

\begin{table*}
\caption{Best present limits on $T_{1/2}$ for decay with one and
two Majorons at 90\% C.L. for modes
with spectral index n = 2, n = 3 and n = 7.}
\begin{tabular*}{\textwidth}{l@{\extracolsep{\fill}}|c|c|c|}
\hline
Isotope ($E_{2\beta}$, keV) & n = 2 & n = 3 &  n = 7  \\
\hline
$^{76}$Ge (2039) & - & $>5.8\cdot10^{21}$ \cite{GUN97} &
$>6.6\cdot10^{21}$ \cite{GUN97} \\
$^{82}$Se (2995) & $>6\cdot10^{21}$ \cite{ARN06} & $>3.1\cdot10^{21}$
\cite{ARN06} & $>5\cdot10^{20}$ \cite{ARN06} \\
$^{96}$Zr (3350) & $>9.9\cdot10^{20}$ \cite{ARG09a}& $>5.8\cdot10^{20}$ \cite{ARG09a} &
$>1.1\cdot10^{20}$ \cite{ARG09a} \\
$^{100}$Mo (3034) & $>1.7\cdot10^{22}$ \cite{ARN06} & $>1\cdot10^{22}$
\cite{ARN06} & $>7\cdot10^{19}$ \cite{ARN06} \\
$^{116}$Cd (2805) & $>1.7\cdot10^{21}$ \cite{DAN03} & $>8\cdot10^{20}$
\cite{DAN03} & $>3.1\cdot10^{19}$ \cite{DAN03} \\
$^{130}$Te (2527) & - & $>9\cdot10^{20}$ \cite{ARN03} & - \\
$^{128}$Te (867) (geochem) & $>1.5\cdot10^{24}$ \cite{MAN91,BAR10a} & $>1.5\cdot10^{24}$ \cite{MAN91,BAR10a} 
& $>1.5\cdot10^{24}$ \cite{MAN91,BAR10a}\\
$^{150}$Nd (3371) & $>5.4\cdot10^{20}$ \cite{ARG09} & $>2.2\cdot10^{20}$
\cite{ARG09} & $>4.7\cdot10^{19}$ \cite{ARG09} \\
\hline
\end{tabular*}
\end{table*}

\section{Tomorrow}

There are more than 20 different propositions for future double beta decay experiments.
Here seven of the most developed and 
promising experiments which can
be realized within the next few years are presented 
(see Table 7). The
estimation of the sensitivity in the experiments is made using
NMEs from \cite{KOR07,KOR07a,SIM08,CAU08,BAR09,CHA08}. In all probability,
they will make it possible to reach the sensitivity for the
neutrino mass at a level of 0.01 to 0.1 eV. 

First phase of GERDA (18 kg of $^{76}$Ge), EXO-200 (200 kg of $^{136}$Xe),
CUORE-0 ($\sim$ 40 kg of natural Te)
and KamLAND-Xe (400 kg of $^{136}$Xe) plan to start data-tacking in 2011.
For this reason I expect occurrance of new, very interesting results in 2011-2012.

\begin{table*}
\caption{Seven most developed and promising projects. 
Sensitivity at 90\% C.L. for three (1-st steps of GERDA and MAJORANA, KamLAND, SNO+) 
five (EXO, SuperNEMO and CUORE) and ten (full-scale GERDA and MAJORANA) 
years of measurements is presented. 
$^{*)}$ For the background 
0.001 keV$^{-1}\cdot kg^{-1} \cdot y^{-1}$; $^{**)}$ for the background 
0.01 keV$^{-1}\cdot kg^{-1} \cdot y^{-1}$. }
\begin{tabular}{|c|c|c|c|c|c|c|}
\hline
Experiment & Isotope & Mass of & Sensitivity & Sensitivity & Status & Start of  \\
& & isotope, kg & $T_{1/2}$, y & $\langle m_{\nu} \rangle$, meV & & data-tacking \\
\hline
CUORE & $^{130}$Te & 200 & $6.5\cdot10^{26}$$^{*)}$ & 20-50 & in progress & $\sim$ 2013 \\ 
\cite{ARNA04,ARNA08} & & & $2.1\cdot10^{26}$$^{**)}$ & 40-90 & \\
GERDA  & $^{76}$Ge & 40 & $2\cdot10^{26}$ & 70-200 & in progress & $\sim$ 2012 \\
\cite{ABT04,JAN09}& & 1000 & $6\cdot10^{27}$ & 10-40 & R\&D & $\sim$ 2015\\ 
MAJORANA & $^{76}$Ge & 30-60 & $(1-2)\cdot10^{26}$ & 70-200 & in progress & $\sim$ 2013 \\
\cite{MAJ03,GUI08}& & 1000 & $6\cdot10^{27}$ & 10-40 & R\&D & $\sim$ 2015\\ 
EXO \cite{DAN00,GOR09} & $^{136}$Xe & 200 & $6.4\cdot10^{25}$ & 100-200 & in progress & $\sim$ 2011 \\
& & 1000 & $8\cdot10^{26}$ & 30-60 & R\&D & $\sim$ 2015\\ 
SuperNEMO & $^{82}$Se & 100-200 & $(1-2)\cdot10^{26}$ & 40-100 & R\&D & $\sim$ 2013-2015\\
\cite{BAR02,BAR04,CHA09} & & & & &\\
KamLAND & $^{136}$Xe & 400 & $4\cdot10^{26}$ & 40-80 & in progress & $\sim$ 2011 \\
\cite{NAK10} & & 1000 & $10^{27}$ & 25-50 & R\&D & $\sim$ 2013-2015\\
SNO+ & $^{150}$Nd & 56 & $4.5\cdot10^{24}$ & 100-300 & in progress & $\sim$ 2012 \\
\cite{MAN10} & &  500 & $3\cdot10^{25}$ & 40-120 &        R\&D  & $\sim$ 2015\\

\hline
\end{tabular}
\end{table*}

\section{Acknowledgements}

This work was supported by Russian Federal Agency for Atomic Energy.

\end{document}